\documentclass[amsmath,amssymb,nofootinbib]{revtex4}

\usepackage{xcolor}
\usepackage{latexsym,comment,verbatim}
\usepackage{amssymb}
\usepackage{amsfonts}
\usepackage{amsmath,color}
\usepackage{graphicx}
\usepackage{bm}
\usepackage{appendix}
\usepackage{pdfpages}
\date{\today}
\begin{document}
\title{GINGER}
\author{Carlo Altucci$^{1,2}$, Francesco Bajardi$^{1,2,3}$, Emilio Barchiesi$^4$, Andrea Basti$^{5,6}$, Nicolò Beverini$^6$, Thomas Braun$^7$, Giorgio Carelli$^6$, Salvatore Capozziello$^{1,2,3}$, Donatella Ciampini$^6$, Fabrizio Davì$^8$, Gaetano De Luca$^9$, Roberto Devoti$^{10}$, Rita  Di Giovambattista$^{10}$, Giuseppe Di Somma$^{2,5}$, Giuseppe Di Stefano$^{11}$, Angela D.V. Di Virgilio$^5$, Daniela Famiani$^{11}$, Alberto Frepoli$^{10}$, Francesco Fuso$^6$,
Ivan Giorgio$^{12}$, Aladino Govoni$^{10}$,  Gaetano Lambiase$^{13,14}$, Enrico Maccioni$^{5,6}$, Paolo Marsili$^6$, Alessia Mercuri$^{11}$, Fabio Morsani$^5$, Antonello Ortolan$^{15}$, Alberto Porzio$^{1,16}$, Matteo Luca Ruggiero$^{17}$, Marco Tallini$^{12}$, Jay Tasson$^{19}$, Emilio Turco $^{4}$ , Raffaele Velotta $^{1,2}$} 

\address{
$^1$ Istituto Nazionale di Fisica Nucleare, Sez. di Napoli,
Compl. Univ. Monte S. Angelo, Edificio G, Via Cinthia, I-80126, Napoli, Italy\\
$^2$ Dipartimento di Fisica “E. Pancini”, Università di Napoli “Federico II”, Napoli, Italy\\
$^3$ Scuola Superiore Meridionale, Largo San Marcellino 10, I-80138, Napoli, Italy\\
$^4$ Dipartimento di Architettura, design e urbanistica, Università degli
Studi di Sassari, Alghero (Italy).\\
$^5$ INFN sezione di Pisa, Pisa, Italy\\
$^6$  Dipartimento di Fisica, Università di Pisa, Pisa, Italy\\
$^7$ Istituto Nazionale di Geofisica e Vulcanologia, Sez. di Roma1, Sede di Arezzo, Italy\\
$^8$ Università Politecnica delle Marche, via Brecce Bianche, 60131 Ancona, Italy\\
$^9$ Istituto Nazionale di Geofisica e Vulcanologia, Sez. ONT, sede di L'Aquila, Italy \\
$^{10}$ Istituto Nazionale di Geofisica e Vulcanologia, Sez. ONT, sede di Roma, Italy \\
$^{11}$ Istituto Nazionale di Geofisica e Vulcanologia, Sez. di Roma1, sede di Roma, Italy \\
$^{12}$ Department of Civil and Construction Engineering, Piazzale Ernesto Pontieri 1, Monteluco di Roio - 67100 L'Aquila, Italy\\
$^{13}$ Dipartimento di Fisica "E. R. Caianiello", Universitá degli Studi di Salerno,
Via Giovanni Paolo II, 132, I-84084, Fisciano (SA), Italy\\
$^{14}$ Istituto Nazionale di Fisica Nucleare (INFN), gruppo collegato di Salerno, Italy \\
$^{15}$ Legnaro INFN National Laboratory, Legnaro\\
$^{16}$ CNR-SPIN, U.O.S. Napoli, Complesso Univ. Monte Sant’Angelo, via Cintia, Napoli, Italy.\\
$^{17}$ Dipartimento di Matematica “G.Peano”, Universit`a degli studi di Torino, Via Carlo Alberto 10,
10123 Torino, Italy.\\
$^{18}$ INFN - LNL , Viale dell’Universit`a 2, 35020 Legnaro (PD), Italy \\
$^{19}$ Physics and Astronomy Department, Carleton College, Northfield, MN 55057, USA
}
\begin{abstract}
    In this paper, we outline the scientific objectives, the experimental layout, and the collaborations envisaged for the GINGER (Gyroscopes IN GEneral Relativity) project. The GINGER project brings together different scientific disciplines aiming at building an array of Ring Laser Gyroscopes (RLGs), exploiting the Sagnac effect, to measure continuously, with sensitivity better than pico-rad/s, large bandwidth (ca. 1 kHz), and high dynamic range, the absolute angular rotation rate of the Earth. In the paper, we address the feasibility of the apparatus with respect to the ambitious specifications above, as well as prove how such an apparatus, which will be able to detect strong Earthquakes, very weak geodetic signals, as well as general relativity effects like Lense-Thirring and De Sitter, will help scientific advancements in Theoretical Physics, Geophysics, and Geodesy, among other scientific fields.

\end{abstract}
\maketitle
\tableofcontents

\section{Introduction}
Sagnac gyroscopes, and in particular Ring Laser Gyroscopes (RLGs), are quite peculiar instruments. At present RLGs are the only instruments able to measure continuously, with sensitivity better than pico-rad/s, large bandwidth ($\sim 1$ kHz), and high dynamic range, the absolute angular rotation rate of the Earth. The latter is a distinctive feature of RLGs, where the high dynamic range is a consequence of the rotation measurement based on instantaneous frequency estimation\cite{uno}. Therefore, a RLG can detect at the same time a strong earthquake, very tiny geodetic signals, and General Relativity (GR) effects, like the Lense-Thirring and de Sitter\cite{Soffel, Kopeikin}. Hence, RLGs have a very wide spectrum of applications ranging from seismology, in the frequency window  $\sim 0.01 \div 30$ Hz, to geodesy and geophysics, in the very low frequency domain ($< 10^{-3}$ Hz), and further down to GR signals lying nearly at zero frequency. \\
In the following the GINGER scientific objectives, the collaborating institutions and the experimental lay-out are
illustrated. At the end, timeline and needs are reported.
\section{Scientific Objectives in Physics}
Light based interferometers have reached an extremely high level of sensitivity, reliability, and robustness. In this perspective, the large interferometers (e.g. LIGO and VIRGO)  are a classic  example. In these devices,  the light beam coming from the light source  splits in two separate beams, injected in two separated paths and recombined to interfere with each other.  The interference fringes bring information on the two different paths. Different pattern topology can be used in view of physical effects and principles under investigation. One of the possibilities is to use a closed polygonal path, for example defined by 4 mirrors located at the vertices of a square, with the two light beams circulating inside the cavity in  clockwise and counter clockwise directions. The interference of the two counter propagating beams brings information on the non reciprocal effects connected to the direction of circulation, since the two spatial paths are identical with each other. If the frame supporting the 4 mirrors of the ring rotates, the two counter propagating beams complete the path at different times, an effect usually called Sagnac effect, after the French physicist Geoge Sagnac, who showed it approximately 100 years ago\cite{Sagnac}.  Accordingly, Sagnac interferometers are a specific class of interferometers, commonly used to measure inertial angular rotation. However, there are other non reciprocal effects related to the propagation of the two light beams that are connected to the space time structure or symmetries, and so related to fundamental physics.\\
The most important instrument in this class is the active Sagnac interferometer, usually referred to as  ``Ring Laser Gyroscope'' (RLG), which allows extraordinarily high resolution measurements of angular velocity. \\
A RLG senses the projection of the angular velocity vector $\vec{\Omega}$ on the   area of the closed polygonal cavity (typically a square). The orientation of this area in space is determined by the area versor $\vec n$. The relationship between the Sagnac frequency $f_s$ and the angular rotation rate $\Omega$ reads 
\begin{equation}
f_s =4\frac{A}{\lambda L} \Omega \cos{\theta} \ , \\
\label{uno}
\end{equation}
where $A$ is the area of the  cavity, $L$ is its perimeter, $\lambda$ is the wavelength of the light, and $\theta$ is the angle between $\vec n$  and  $\vec{\Omega}$.
The GINGER (Gyroscopes IN GEneral Relativity) project is based on an RLG array, with each RLG attached to the Earth crust. From now on, it is therefore assumed that $\Omega = \Omega_T$, the Earth angular rotation $\Omega\oplus$ plus all other non reciprocal effects between the two counter-propagating beams. Each RLG of the array measures a different projection of  $\vec{\Omega}_T$. 
The main objective of the instrument is to reconstruct the total angular velocity vector $\vec{\Omega}_T$, which contains the kinematic term $\vec{\Omega}_\oplus$, contributions due to gravity (e.g. Lense-Thirring and de Sitter effects of GR), effects of  Lorentz violation (if any), and local contributions from geophysics and geodesy. The International Earth Rotation System (IERS) measures and makes available the kinematic term with very high accuracy, and so gravitational theories can be tested by comparing the independent measurements of RLGs and IERS. The effectiveness of GINGER for fundamental physics investigation depends on its sensitivity, which quite often is expressed as the relative precision in the  Earth angular rotation rate measurement; it can be said that 1 part in $10^{9}$ is a meaningful target for fundamental physics, and it is certainly feasible since it has been already obtained\cite{A2, DiVirgilio2022}. We expect GINGER to reach the 1 part in $10^{11}$ relative precision in a four year time. Further improvements by 1 or 2 orders of magnitude seems at present feasible, but it is necessary to proceed step by step. We remark that 1 part in $10^{13}$  is equivalent to test the Lense-Thirring on Earth at the $1\%$ level.

In this paper, we outline the scientific objectives, the experimental layout, and the collaborations
envisaged for the GINGER (Gyroscopes IN GEneral Relativity) project. The GINGER project
brings together different scientific disciplines aiming at building an array of Ring Laser Gyroscopes
(RLGs), exploiting the Sagnac effect, to measure continuously, with sensitivity better than pico-
rad/s, large bandwidth (ca. 1 kHz), and high dynamic range, the absolute angular rotation rate of
the Earth. In the paper, we address the feasibility of the apparatus with respect to the ambitious
specifications above, as well as prove how such an apparatus, which will be able to detect strong
Earthquakes, very weak geodetic signals, as well as general relativity effects like Lense-Thirring and
De Sitter, will help scientific advancements in Theoretical Physics, Geophysics, and Geodesy, among
other scientific fields.

\subsection{Testing Theories of Gravity and  Lorentz invariance}
The debate on gravitational theories, to extend or modify General Relativity, is very active for the issues related to ultra-violet  and infra-red behavior of Einstein's theory. The first is connected to the Quantum Gravity problem, the other one to Dark Matter and Dark Energy governing large scale structure and cosmology. Up to now, no final theory, capable of explaining gravitational interaction at any scale, has been formulated, and it is important to say that discriminating between different theories certainly requires a sensitivity breakthrough. Earth based experiments can be in principle advantageous allowing a thorough analysis of the systematics, little modelling of external perturbations, and the feasibility of upgrades. The GINGER project, an array of Sagnac gyroscopes, can be used for Earth based gravity measurements, in particular to constrain parameters of gravity  theories, like  scalar-tensor  or Horava-Lifshitz gravity, by considering their post-Newtonian limits matched with experimental results.\\
The relevant effects have been measured by the Gravity Probe B space experiments ($10\%$ level) and by Lageos and Lares,  reaching the  $4\%$ level. Gravity Probe B has finished its operational life, while Lares is still providing data, and Lares II will be launched soon. Tests are based on reconstructing the trajectories of geodetic satellites, i.e. simple mechanical objects to minimise the effect of the different drags affecting the trajectories, and combining laser ranging data coming from different stations. The Lense-Thirring test, using the reconstruction of the nodes of the trajectories, requires the independent map of the Earth gravity fields, independently measured, and the final result at the end is limited by the accuracy  of the map. It is necessary to accurately model the zonal tides, since their influence on the nodes of the geodetic satellites is larger than the Lense Thirring itself.\\
 The test provided by GINGER, which, being Earth-based, provides the measurement at a fixed latitude, is very different. The Lense-Thirring, relying on the Sagnac effect, does not require to combine data coming from different places or to use an accurate map of the Earth gravitational field.\\ 
Recently, it has been pointed out by Jay Tasson of Carleton University that RLG can effectively contribute to the Lorentz Violation quest. In this respect Jay Tasson has shown that a level of sensitivity of 1 part in $10^9$ for the Earth rotation rate would provide interesting measurements of two Lorentz-violating terms in the framework of the Standard-Model Extension.  In one case, sensitivities that are competitive with recent laboratory, and perhaps solar system,  tests would result. 
For the other term, measurements competitive with the best existing limits, which currently come from radio pulsar studies, would result\cite{A1}. From an experimental point of view the Lorentz Violation test is done using high frequency harmonics. 
\begin{figure}[h]
    \centering
 \includegraphics[scale=0.25]{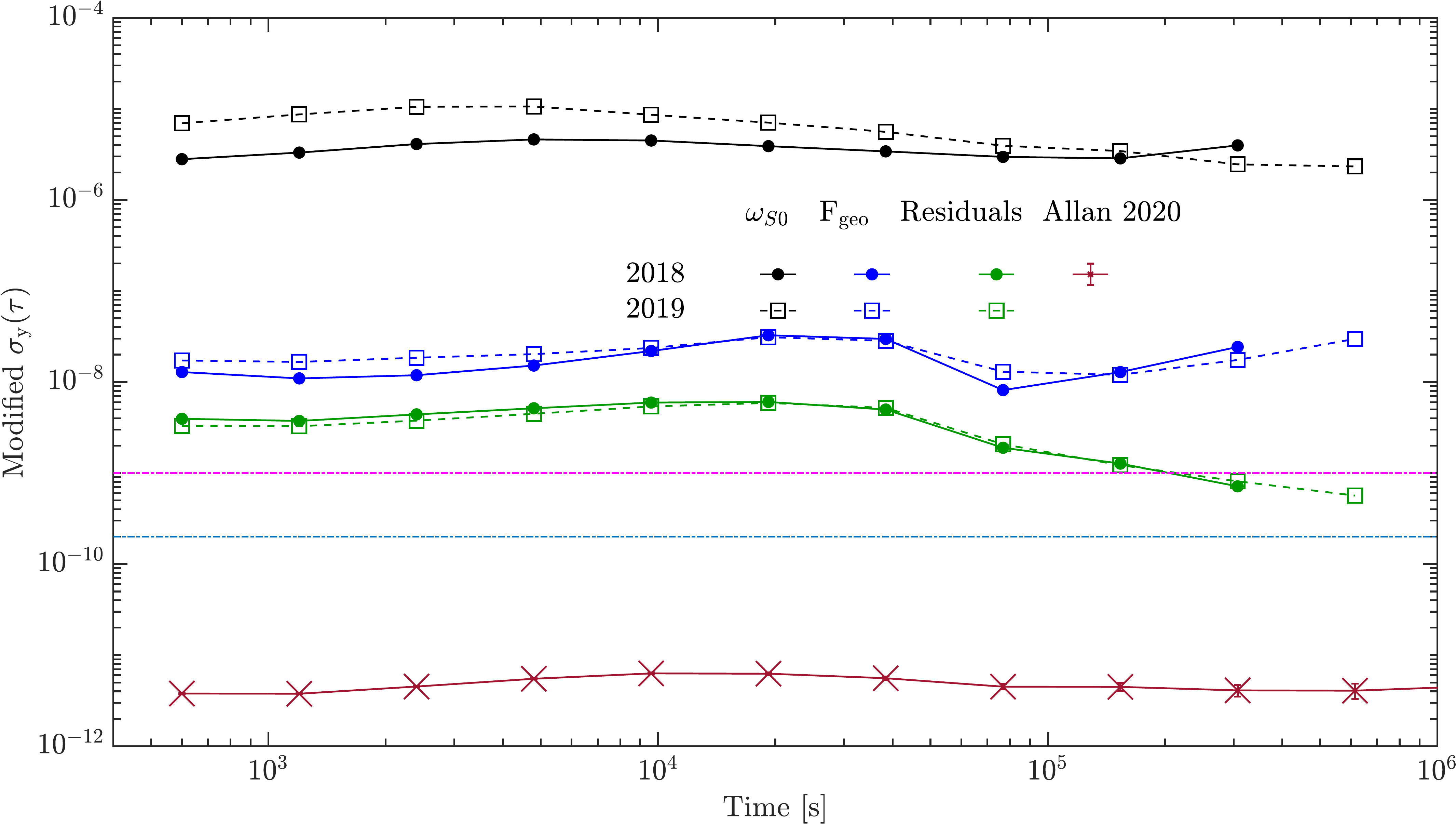}
    \caption{ Modified Allan Deviation $\sigma_y(\tau)$ calculated for $\omega_{s0}$ (the first level of reconstruction of the Sagnac frequency), for $F_{geo}$, and for the residuals defined as in the text. 
    Two different data sets acquired by GINGERINO are used taken in different years: both 2019 (dotted line) and 2018 (solid line) are shown.
    Error bars are not visible in this scale. The two horizontal dot-dashed lines indicate the two levels meaningful for Lorentz Violation and Lense-Thirring tests, the top one being the first one. The lowest curve is the Allan deviation of the residuals evaluated improving the analysis procedure utilised for the 2018 and 2019 data sets, showing that the sensitivity is close to 0.1 frad/s, paving the way to push the Lense-Thirring test down to 0.1$\%$ level\cite{ventidue}}
    \label{fig:MADl}
\end{figure}

\begin{figure}[h]
    \centering
 \includegraphics[scale=0.25]{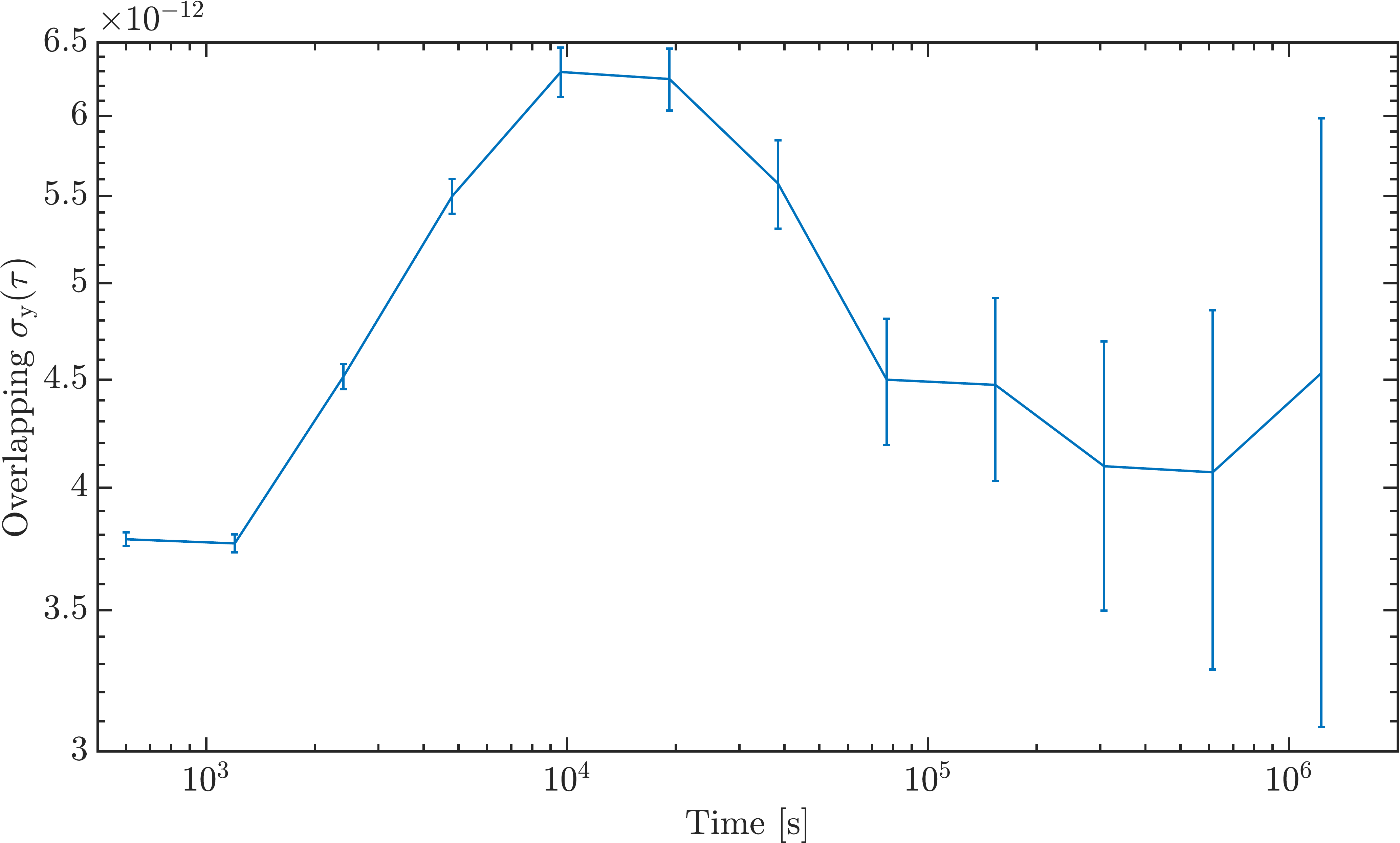}
    \caption{Detail of Overlapping Allan Deviation $\sigma_y(\tau)$ calculated for the residuals of the analysis.}
    \label{fig:MADl1}
\end{figure}

It can be interesting to point out that the analysis of the GINGERINO data is compatible  with 0.1 frad/s limiting sensitivity\cite{ventidue}, GINGERINO is a single 4 m-side RLG located underground in the Gran Sasso National Laboratory of INFN (LNGS). The RLG runs unattended since several years, providing data with a good duty cycle, better than 80\%. A complete set of analysis procedures has been developed in order to take into account all effects relating with the inherently nonlinear laser dynamics\cite{AA26, AA28}.

Figure \ref{fig:MADl} reports the Modified Allan Deviation of the residuals evaluated subtracting $F_{IERS}$ and $F_{geo}$; the first one is the expected signal in the gyroscope due to the wobbles, polar motion, and $\Omega_\oplus$ variations (evaluated using IERS data), the latter is the signal reconstructed using the data, evaluated with a linear regression taking into account laser dynamics and the environmental monitors.  The residuals provide a direct estimate of the sensitivity limit of the apparatus.\cite{A2}
The two horizontal lines in Figures \ref{fig:MADl} and \ref{fig:MADl1} show the boundary expected to be meaningful for the Lorentz Violation and the Lense-Thirring effect.

Figure \ref{fig:MADl} shows that already the target of 1 part $10^9$ is meaningful. The target of $1\%$ of the Lense-Thirring effect is equivalent to reach 1 part in $10^{12}$, which is feasible extending the measurement time and keeping under control systematics in general. Regarding Lorentz Violation, GINGERINO has already shown that 1 part in $10^9$ is feasible, but unfortunately its orientation is not optimal for this purpose.
It is important to remind that the Earth frame dragging, or Lense-Thirring, has been already measured with space based experiments\cite{due, quattro, cinque, sei, sette}, the main difference between GINGER and the space experiments is that the measurement is at a given latitude, not averaged and it does not require the independent map of the gravity field.
\subsection{Fundamental Physics Issues}
As a gyroscope, the goal of GINGER is to measure the Earth angular rotation vector (modulus and direction) referenced
to the local inertial frame, with a relative precision better than 1 part in $10^{10}$, that is to achieve a rotational sensitivity
of at least a few frad/s. As light is an intrinsically relativistic probe of classical and quantum spacetimes, the
expected GINGER sensitivity translates into several scientific goals:
\break
\begin{enumerate}
    \item To detect effects due to the curvature of space-time around the Earth (de Sitter effect)
and to the spinning of the Earth mass (dragging of the inertial frames or Lense-Thirring effect). This measurement
requires the comparison between the IERS data of
the Earth rotation vector and the corresponding GINGER data.
\item To test General Relativity extension/modifications by improving accuracy and reliability of data. Some of the expected
measurements can indeed be considered as upper limits; accordingly, any improvement in sensitivity and accuracy
can pave the way for further theoretical insights in the theories of gravitation beyond General Relativity.
\item To test Lorentz violations as described by Standard Model Extension (SME). In fact, it has been pointed out
that SME terms with dimension d=4 and d=5 can also break the symmetry for the counter-propagating beams in
a RLG, and that GINGER can effectively contribute to the Lorentz Violation quest. Also in this case the signal can
be inferred via comparison of GINGER with IERS data. It is important to say that this test is based on the observation at fixed frequency, not on a DC level, so high accuracy is not required.
\item To investigate whether fluctuations due to the graininess of spacetime can potentially lead to observable
signatures in high frequency RLG spectra. Intuitively, the natural length and time scales associated with the
quantum nature of spacetime is the Planck length, and its fluctuations give rise to a white noise that can be
investigated using the frequency comb with harmonics at integer multiples of the RLG free spectral range.
\item Gravitational waves can excite the normal modes of the Earth. Detection of such signals seems in principle feasible,
provided that the sensitivity is higher than $10^{-16}$ rad/s.

\item In a  recent proposal Marletto and Vedral \cite{venticinque} have pointed out the possibility to explore, via (a quantum version of the) Sagnac
interferometer, the quantum nature of gravity. This analysis
assumes the validity of the equivalence principle formulated in its
quantum version.

\item Sagnac corrections to the time delay has been derived in the context of
Horava-Lifshitz gravity, which is
a power-counting renormalizable theory, and therefore can be considered as
a candidate for the UV completion of GR.

\item  The unification  of GR and quantum mechanics principles
is a long-standing issue of contemporary physics. Experimental techniques
in quantum optics have only recently reached the precision required for
the investigation of quantum systems under the influence of non-inertial
motion, such as being held at rest in gravitational fields or subjected to
uniform accelerations. In this perspective, it will be interesting to
explore the entanglement phenomena or quantum mechanics tests in
non inertial reference frames.
\end{enumerate}

\subsection{Experimental measurement of the Lense-Thirring Effect}\label{sec:intro}
Table \ref{table1} reports the experiments, past and future, aiming at the Lense Thirring measurement. In the notes, main limitations and general status are indicated.
\begin{table}
\begin{tabular}{c|c|c|c|c}
Year &  Experiment and Location & Uncertainty &References & Notes  \\\hline
2011 & Gravity Probe B - Space, Earth gravity field  &  20\% &\cite{everitt2011gravity} & \textbf{a}  \\\hline 
2019 & LARES, LAGEOS,  LAGEOS 2 - Space, Earth gravity field  &  2\% &\cite{ciufolini2019improved} & \textbf{b}  \\\hline 
2019-2020 & LARES, LAGEOS,  LAGEOS 2 - Space, Earth gravity field& 1\%& \cite{lucchesi2019general,lucchesi20201} & \textbf{c}  \\\hline
2020 & Pulsar Timing - Space, Binary System PSR J1141- 6545 & ???& \cite{krishnan2020lense,iorio2020comment} & \textbf{d}  \\\hline
1999 & HYPER: Cold Atoms Interferometry - Space, Earth gravity field  & 1\% (planned) & \cite{jentsch2004hyper,angonin2004gravitational,tino2021testing} & \textbf{e}  \\\hline
2018 & Pulsar Timing - Space, Binary System PSR J0737-3039A/B & 10\% (planned) & \cite{kehl2018future} & \textbf{f}  \\\hline
2020 & Pulsar Timing - Space, Binary System PSR J0737-3039A/B & 7\% (planned) & \cite{hu2020constraining} & \textbf{g}  \\\hline
2021 & Gravity Probe Spin, ferromagnetic gyroscope - Space, Earth gravity field  &  (proposal) & \cite{fadeev2021gravity} & \textbf{h}  \\\hline
\end{tabular}
\vspace{1cm}
\begin{raggedright}

\noindent \textbf{a}: large torques on the rotors were caused by  electrostatic interactions between surface imperfections (patch effect).\\

\noindent \textbf{b}: the estimated systematic error is due to modelling errors in the orbital perturbations, mainly due to the errors in the Earths gravity field determination.\\

\noindent \textbf{c}: improvement in the modelling of the even zonal harmonics of the Earth gravitational field, using GRACE coefficients in the time interval of analyses  (6.5 years); difficult to disentangle gravitational periodic perturbations from non gravitational ones.\\

\noindent \textbf{d}: the temporal evolution of the orbital inclination of  is caused by a combination of a Newtonian quadrupole moment and Lense-Thirring precession of the orbit, but the physical and orbital systems parameters are not known independently; moreover, the uncertainty due to the quadrupole mass moment is missing.\\

\noindent \textbf{e}: after an assessment and an industrial study, ESA decided not to continue the development of this mission because the technology readiness level was considered too low.\\

\noindent \textbf{f}: this measurement is expected to be performed with SKA. \\

\noindent \textbf{g}: this measurement is expected to be performed with SKA and MeerKAT. \\

\noindent \textbf{h}:  the expected uncertainty in the precession frequency (with the lower limit $10^{-15}$ s$^{-1}$) is determined by background gas collisions and by magnetic field drifts.

\end{raggedright}
\caption{Summary of experiments aimed at Lense-Thirring measurements.}\label{table1}
\end{table}

\color{black}

\subsection{Testing the  PPN parameters}
Gravity Probe B has measured the de Sitter (geodetic) precession at $0.3\%$ ($\gamma$).
The Parametrized Post Newtonian (PPN) parameters have not been evaluated by Lares and Lageos and Pulsars data.
Hyper-space experiment with atomic clocks has been proposed to measure the PPN 
parameters at the level of
$10^{-16}$rad/s, but this part of the proposal has not been approved.
Gravity Probe B  proposes the measurement of de Sitter and spin  precessions, but there are no specifications regarding the PPN.
In principle, GINGER would be sensitive to parameters $\alpha$, $\beta$, $\gamma$ of the Eddington parametrization of gravity \cite{Will:2014kxa}, which point out possible deviations with respect to GR. In fact, if they are rigorously $\alpha= \beta= \gamma =1$, GR is restored. Specifically, $\alpha =1$ maens that the value of the gravitational constant is the Newtonian one, $\beta$ indicates ``non-linearities'' present in the gravitational theory, while $\gamma$ is related to the spatial curvature produced by a unitary mass at rest. In other words, these parameters  are related to the potentials which are ``entries'' of the space-time metric.
By a rapid inspection of Tab.~\ref{table1}, it appears that any extension/modification of GR implies that some of the PPN parameters differs from the rigorous value 1. In summary, besides spatial based experiments, GINGER could be a powerful tool to  finely test gravity theories (see \cite{Capozziello:2021goa} for details).

\subsection{Extensions and modifications of General Relativity}

Several alternatives to GR are considered in  literature with the aim to solve the shortcomings
suffered by Einstein’s theory at any scale of energy ranging from ultraviolet to infrared scales. The main issues to invoke extensions or modifications of GR 
are related to the lack of a final and self-consistent theory of Quantum Gravity and to the fact that, up to now, there are no  answers, at the fundamental level, to address 
the problems of dark energy and dark matter  (see \cite{Capozziello:2011et, Capozziello:2007ec,Capozziello:2021krv,Bamba:2012cp} for comprehensive reviews).
A pictorial view to show how these extensions/modifications stem out from  GR is reported in Fig.~\ref{fig:GR_interconnections}. It is worth noticing that most of them lead to corrections to the Newtonian potentials in the weak-field limit. Some of these corrections  are summarized in the following Table:
\begin{figure}
    \centering
    \includegraphics[scale=0.5]{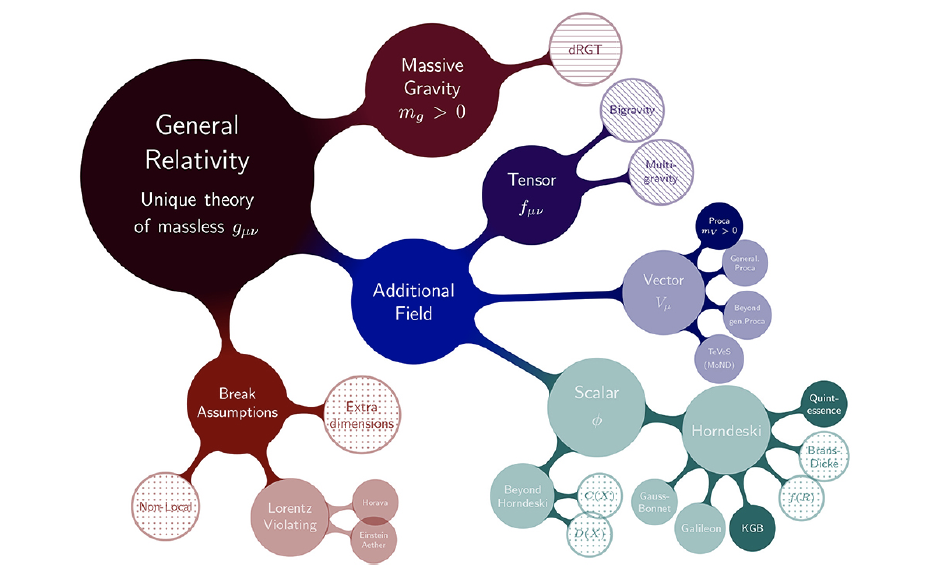}
    \caption{Extensions of General Relativity.}
    \label{fig:GR_interconnections}
\end{figure}
\newpage 
\resizebox{16.6cm}{!}{
\begin{centering}
\begin{tabular}{c|c|cl}
\hline
  Modified  Gravity Model  & Corrected potential & Yukawa parameters \\
\hline &&&
\\
 $f(R)$ & $\begin{array}{ll} \Phi(r)=-\frac{G_N M}{r}\biggl[1+\alpha\,e^{-m_R r}\biggl] \\ \displaystyle \textbf{A} = \frac{G_N}{r^3} \, \textbf{r} \times \textbf{J} \end{array}$ & $\begin{array}{ll}{m_R}^2\,=\,-\frac{f_R(0)}{6f_{RR}(0)}\end{array}$ \\ \\
\hline
 & & &\\ 
$f(R,\Box R)= R+ a_0R^2+a_1R\Box R$ & $\begin{array}{ll} \Phi(r)=-\frac{G_NM}{r}\left(1+c_0e^{(-r/l_0)}+c_1e^{(-r/l_1)}\right) \\ \displaystyle \textbf{A} = \frac{G_N}{r^3} \, \textbf{r} \times \textbf{J} \end{array}$ & $\begin{array}{ll}{c_{0,1}}\,=\,\frac{1}{6}\mp\frac{a_0}{2\sqrt{9a_0^2+6a_1}}\\\\  l_{0,1}\,=\,\sqrt{-3a_0\pm\sqrt{9a_0^2+6a_1}}\end{array}$ \\
\hline &&&
\\
$f(R,\Box R,..\Box^k R)= R+\Sigma_{k=0}^p a_kR\Box^k R$ & $\begin{array}{ll} \Phi(r)=-\frac{G_NM}{r}\left(1+\Sigma_{k=0}^{p}c_i\exp(-r/l_i)\right) \\ \displaystyle \textbf{A} = \frac{G_N}{r^3} \, \textbf{r} \times \textbf{J} \end{array}$ & $c_i\,,\;l_i$ are functions of $a_k$. See \cite{Quandt:1990gc}. \\ \\
\hline
 & & &\\
   $f(R,\,R_{\alpha\beta}R^{\alpha\beta})$ & $\begin{array}{ll}\Phi(r)=-\frac{G_NM}{r}\biggl[1+\frac{1}{3}\,e^{-m_R r}-\frac{4}{3}\,e^{-m_Y r}\biggl] \\ \displaystyle \textbf{A} = \frac{G_N}{r^3} \, \textbf{r} \times \textbf{J} \end{array}$ & $\begin{array}{ll}{m_R}^2\,=\,-\frac{1}{3f_{RR}(0)+2f_Y(0)} \\\\{m_Y}^2\,=\,\frac{1}{f_Y(0)}\,,\, Y=R_{\mu\nu}R^{\mu\nu}
\end{array}$ \\ \\
\hline  
& & & \\  
$f(R,\,\cal{G})$\,, \; ${\cal G}= R^2-4R_{\mu\nu}R^{\mu\nu}+ R_{\alpha\beta\mu\nu}R^{\alpha\beta\mu\nu}$& $\begin{array}{ll}\Phi(r)=-\frac{G_NM}{r}\biggl[1+\frac{1}{3}\,e^{-m_1 r}-\frac{4}{3}\,e^{-m_2 r}\biggl] \\ \displaystyle \textbf{A} = \frac{G_N}{r^3} \, \textbf{r} \times \textbf{J} \end{array}$  & $\begin{array}{ll}{m_1}^2\,=\,-\frac{1}{3f_{RR}(0)+2f_Y(0)+2f_Z(0)} 
\\\\{m_2}^2\,=\,\frac{1}{f_Y(0)+4f_Z(0)}\,,\,Z=R_{\alpha\beta\mu\nu}R^{\alpha\beta\mu\nu}
\end{array}$\\ \\
\hline  & & & 
\\  
$f(R,\,\phi)+\omega(\phi)\phi_{;\alpha}\phi^{;\alpha}$ &
$\begin{array}{ll}\Phi(r)=-\frac{G_NM}{r}\biggl[1+g(\xi,\eta)\,e^{-m_R\tilde{k}_R r}+\\\\\qquad\qquad+[1/3-g(\xi,\eta)]\,e^{-m_R\tilde{k}_\phi r}\biggr] \\ \\  \displaystyle \textbf{A} = \frac{G_N}{r^3} \, \textbf{r} \times \textbf{J}\end{array}$
&
$\begin{array}{ll}{m_R}^2\,=\,-\frac{1}{3f_{RR}(0,\phi^{(0)})}\\\\{m_\phi}^2\,=\,-\frac{f_{\phi\phi}(0,\phi^{(0)})}{2\omega(\phi^{(0)})}\\\\\xi\,=\,\frac{3{f_{R\phi}(0,\phi^{(0)})}^2}{2\omega(\phi^{(0)})}\\\\\eta\,=\,\frac{m_\phi}{m_R}\\\\g(\xi,\,\eta)\,=\,\frac{1-\eta^2+\xi+\sqrt{\eta^4+(\xi-1)^2-2\eta^2(\xi+1)}}{6\sqrt{\eta^4+(\xi-1)^2-2\eta^2(\xi+1)}}\\\\{\tilde{k}_{R,\phi}}^2\,=\,\frac{1-\xi+\eta^2\pm\sqrt{(1-\xi+\eta^2)^2-4\eta^2}}{2}
\end{array}$ \\
\hline & & & 
\\  
$f(R,\,R_{\alpha\beta}R^{\alpha\beta},\phi)+\omega(\phi)\phi_{;\alpha}\phi^{;\alpha}$
&
$\begin{array}{ll}\Phi(r)=-\frac{G_NM}{r}\biggl[1+g(\xi,\eta)\,e^{-m_R\tilde{k}_R\,r}+\\\\\,\,\,\,+[1/3-g(\xi,\eta)]\,e^{-m_R\tilde{k}_\phi\,r}-\frac{4}{3}\,e^{-m_Y r}\biggr] \\ \\ \\  \\ \displaystyle \textbf{A} = \frac{G_N}{r^3}\left[1 - \left( 1+ m_Y r  \right)e^{- m_Y r} \right]\textbf{r} \times \textbf{J}\end{array}$ 
&
$\begin{array}{ll}{m_R}^2\,=\,-\frac{1}{3f_{RR}(0,0,\phi^{(0)})+2f_Y(0,0,\phi^{(0)})}\\\\{m_Y}^2\,=\,\frac{1}{f_Y(0,0,\phi^{(0)})}\\\\{m_\phi}^2\,=\,-\frac{f_{\phi\phi}(0,0,\phi^{(0)})}{2\omega(\phi^{(0)})}\\\\\xi\,=\,\frac{3{f_{R\phi}(0,0,\phi^{(0)})}^2}{2\omega(\phi^{(0)})}\\\\\eta\,=\,\frac{m_\phi}{m_R}\\\\g(\xi,\,\eta)\,=\,\frac{1-\eta^2+\xi+\sqrt{\eta^4+(\xi-1)^2-2\eta^2(\xi+1)}}{6\sqrt{\eta^4+(\xi-1)^2-2\eta^2(\xi+1)}}\\\\{\tilde{k}_{R,\phi}}^2\,=\,\frac{1-\xi+\eta^2\pm\sqrt{(1-\xi+\eta^2)^2-4\eta^2}}{2}
\end{array}$ \\
\hline 
\end{tabular}
\end{centering}
} 
\newpage
The table shows the potential and the associated parameters for several extensions of GR, with $f(0)$ being the first derivative of $f(R)$ with respect to the scalar curvature $R$, evaluated at $R = 0$. It is worth noticing that, in the weak field limit, all these theories exhibit  combinations of Yukawa terms ruled by some sensitive parameters always associated with some effective mass \cite{CapFar}.
It is worth noticing that predictions of GR have been confirmed with extremely high accuracy \cite{sei}, however, the challenge of future 
experiments is to search for deviations from GR, and, in particular, from Newtonian gravity (usually denoted as fifth force \cite{Fishbach}).
In particular, the above table clearly shows that several modifications of  GR can be parametrized, in the weak field limit, as combinations of Yukawa-like corrections to the Newtonian potential. This means that fixing the compatibility ranges  of Yukawa-like parameters can be a fundamental tool to discriminate among viable gravitational models.

\begin{figure}
    \centering
    \includegraphics[scale = 0.5]{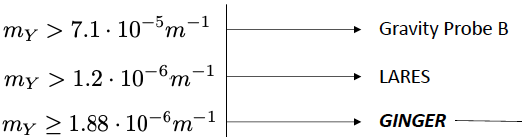}
    \caption{ Comparison of the parameter $m_Y$ of one of the models in the above table between GINGER and other measurements reported in \cite{Capozziello:2014mea}.}
    \label{confr}
\end{figure}

\subsection*{The case of Yukawa-like corrections}

Specifically, considering the definition of  circular velocity $v_c(r) = \sqrt{ r \frac{\partial{\Phi(r)}}{\partial(r)}}$, the 
GINGER experiment can be used to constrain the effective potentials of modified theories of gravity and, consequently, the corresponding action.    Once the effective potential is constrained, the gravitational model is selected. Accordingly, the free parameter values can be compared with those selected by other experiments, in particular those from satellites \cite{Capozziello:2014mea}. 

Another possibility is to consider Gyroscopic and Lense-Thirring effects and constrain the vector potential coming from the PPN limit, in the table indicated with $\mathbf{A}$.  Figure \ref{confr} reports a comparison of $m_Y$ evaluated in different experiments \cite{Capozziello:2014mea} and the GINGER expectation.

Furthermore, the techniques of optical interferometry provide very important results in the framework of gravitation, such as the discovery of gravitational waves  by LIGO/VIRGO experiment, as well as the detection of the gravitomagnetic effect \cite{Capozziello:2014mea}, which can be considered among the corrections to the Newtonian gravitational potential (see the $\mathbf{A}$-terms in the table above).

\begin{figure}
\centering
\includegraphics[scale=0.7]{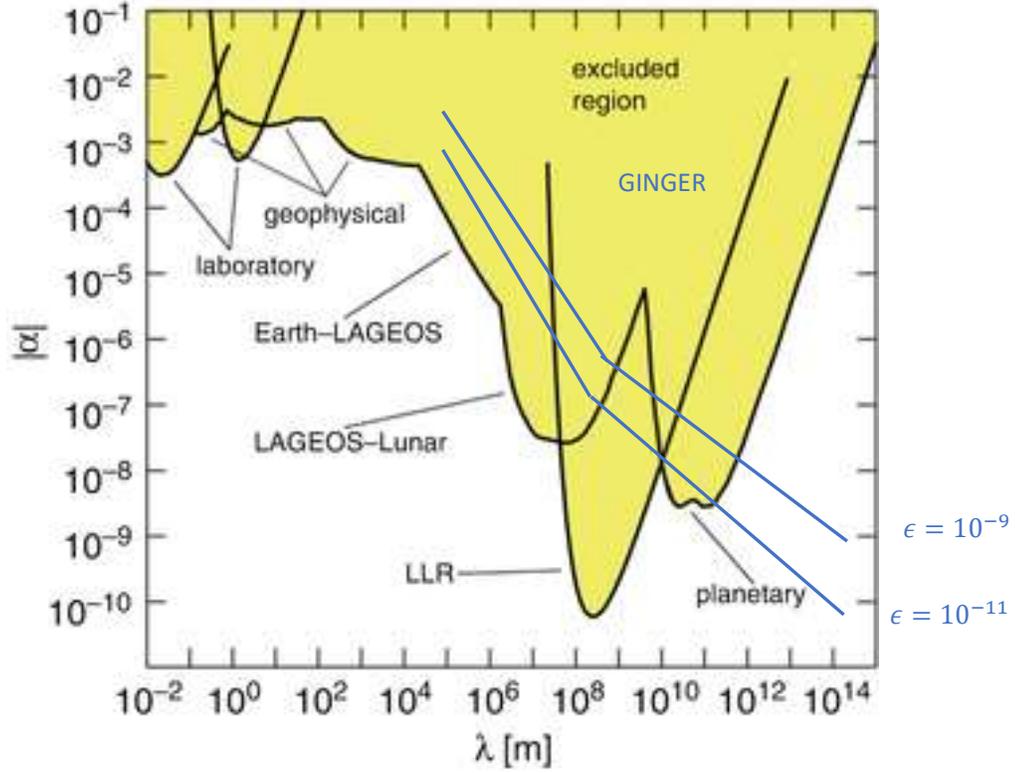}
\caption{The regions of parameter space $\alpha$ vs $\lambda$ accessed by different experiments is shown in the plot. We also report the exclusion region expected with the GINGER experiment, owing to its sensitivity.}
\label{plot1}
\end{figure}

With these considerations in mind, let us take into account a general form of the Yukawa potential \cite{Capozziello:2009vr}
\begin{equation}\label{Yukawapot}
U(r) = - \frac{GM}{r}
\Bigl\{
1 + \alpha\,  e^{-r/\lambda}
\Bigr\}\,. 
\end{equation}
The GINGER data can yield to a validity range for $\alpha$ and $\lambda$.
For terrestrial experiments, it is convenient to make the following approximation:
$r = R + z$, with $z \ll R$, so that (to first order in $z/R$) the potential (\ref{Yukawapot}) reads
\begin{equation}\label{YukawapotApprx}
U(r) = - \frac{G_{\infty}M}{r}
\Bigl\{
1 - \frac{z}{R} +
\alpha \, e^{-R/\lambda} \, \Bigl(1 -  \frac{R
+\lambda}{\lambda}\frac{z}{R}\Bigr)
\Bigr\}\,. 
\end{equation}
The analysis of Sagnac Interferometry in Yukawa potential gives 
(assuming $\vert\alpha\frac{R +\lambda}{\lambda}\exp\{-R/\lambda\}\vert <1$) \cite{Camacho:2003cj}
\begin{equation}\label{Yukphasediff}
\Delta\theta^{(Yukawa)} = -\Delta\theta^{(Newton)}\left[\alpha\frac{R +
\lambda}{\lambda} \, e^{-R/\lambda} \right]\;,
\end{equation}
where 
\begin{equation}\label{Newphasediff}
\Delta\theta^{(Newton)} = \frac{8\pi^2 a^2\nu\Omega}{(c^2 -
a^2\Omega^2)(1 - a\frac{GM}{c^2R^2})}
\frac{1}{
\Bigl\{1 - \cos\Bigl(\frac{2\pi\Omega a}{c}[1 +
\frac{a\Omega}{c}]^{-1} \Bigr)\Bigr\}}
\end{equation}
is the phase shift in Newtonian potential. Here $\Omega$ is the angular velocity of the
interferometer (it rotates in the clockwise direction), $a$ its radius, and $\nu$ the beam frequency
(here the gravitational red-shift has been taken into account). 
From Eq. (\ref{Yukphasediff}) we get
 \begin{equation}\label{sensibility}
|\epsilon| \equiv  \Big|\frac{\Delta\theta^{(Yukawa)}}{\Delta\theta^{(Newton)}} \Big| = \alpha\frac{R +
\lambda}{\lambda}\exp\{-R/\lambda\}\,.
 \end{equation}
Therefore, for a given sensitivity $\epsilon$, a certain region of the  $\{\alpha, \lambda\}$ parameter space can be explored. 
Results are reported in Fig.~\ref{plot1}, where $\epsilon \lesssim 10^{-9}$ and $\epsilon \lesssim 10^{-11}$ are assumed as sensitivities provided by the GINGER experiment. The figure shows as a comparison the parameter space accessible by data coming from other experiments.\\ 
In conclusion, the Sagnac interferometry is a promising candidate to detect gravitomagnetic effects and to offer new possibilities to probe the non--Newtonian gravity. Moreover,  the improvements in  sensitivity/resolution enabled by GINGER will allow to explore new parameter regions and to set more stringent bounds on gravitational theories beyond GR. 

\newpage
\color{black}
\subsection{Testing the Lorentz violation}
\begin{figure}
    \centering
    \includegraphics[scale = 0.5]{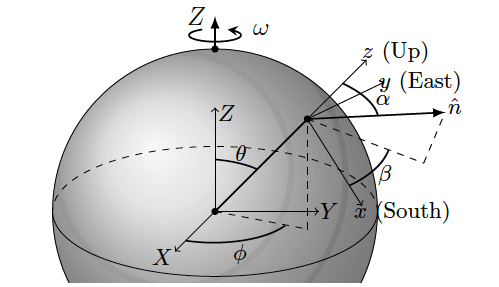}
    \caption{Location of the laboratory in shifted Sun-centered frame axes (X, Y, Z) and
orientation of the normal vector of the gyroscope, area versor \textbf{n}, relative to laboratory coordinates
(x, y, z).}
    \label{LV_ref}
\end{figure}

For a basic discussion about Lorentz Violation (LV) it is necessary to  take into account that the gyroscope is attached in one point on the Earth surface, as sketched in  Fig.~\ref{LV_ref}, please see ref. \cite{A1} for more details.

It has been shown that, in the SME formalism, GINGER data can be used for LV investigation in the gravity sector\cite{Kosteleky}. 
The relevant combinations are $\overline{s}^{TJ}$ and $K_{LKLM}$.
Assuming a resolution of 1 part in $10^9$ of the Earth rotation  rate, the first parameter is relevant for the $d = 4$ dimension and its evaluation is competitive with existing laboratory based measurements
($\overline{s}^{TY} \approx
 \epsilon\frac{\Omega_\oplus R^2}{GM} \approx 10^{-6}$,
 where $\epsilon = 10^{-9}$ indicates the accuracy of the Earth rotation rate measurement). 
It is important to note that the angle $\alpha$ in Fig.~\ref{LV_ref} is included in between the local vertical and the area vector. The LV signals are proportional to $\sin{\alpha}$; as a consequence, existing RLGs, as GINGERINO, are not useful to the purpose, because $\alpha = 0$. 

For evaluating LV combinations, a large number of different experiments are used, in particular pertaining to the gravity sector: cosmic rays,  atomic interferometers,
LLR, 
LIGO/VIRGO, 
superconducting gravimeter arrays,
ultra cold neutrons,
pulsars, 
VLBI, 
bynary pulsars, 
geodetic effects,
Gravity probe B,
short range gravity, 
perihelion precession,
Lageos,
solar spin,
gravitational red shift, torsion pendulum, and tungsten oscillators. For the combination relevant for GINGER the situation is at present the following: for $d=4$ the best limit comes from the array of gravimeters,  
see
Superconducting-Gravimeter Tests of Local Lorentz Invariance\cite{Flowers}, 
while for 
$d=5$ it comes from astrophysical observations, see Testing velocity-dependent CPT-violating gravitational forces with radio pulsars\cite{Shao}.

\section{Geophysics and Geodesy}

Due to the very high sensitivity of the apparatus, which records any tiny motion of the Earth crust (global and local motions), GINGER will have a big impact on both geodesy and geophysics\cite{AA7,AA8,uno,G2,G3,G4,GA4}: Among signals of interest in geodesy, the polar motion, mainly composed of daily variations, and the Annual and Chandler wobbles have been already observed by monolithic RLG. The Earth angular rotation rate is routinely and constantly measured by IERS on a daily basis and the combination of IERS with GINGER data will enable measuring its sub-daily fluctuations.
Moreover, variations of the Length of the Day (LOD) affect the Earth angular rotation rate, and it will be possible to measure them with unprecedented precision and on a sub-hourly base. \\
On the geophysics side, theoreticians have pointed out that, to fully characterize seismic source, in addition to the translational motion also the rotational motion must be measured.
The study of rotational motions in seismology and its implications in related fields such as, seismology, seismic monitoring, earthquake engineering, strong-motion seismology, tectonics, volcanic seismology, seismic sources, etc., relies on the accuracy of the sensors. 

\subsection{Seismology}
Using the new rotational observables provided by GINGER, we expect significant advances in understanding the origin of the ocean-generated noise field, and in solving seismic inverse problems for localizing earthquake sources. 
Remarkably, the Gran Sasso region is one of the most important active seismic areas, and the presence of GINGER will contribute to understanding and studying these problems with an unprecedented degree of detail.\\
During the transit of seismic waves the Earth ground is not only translating but it also rotates. Traditionally, seismologists could only measure translations along three cardinal axes, but earthquakes also generates tilt motion which rotates the ground. This fact was predicted by the linear elasticity theory, but it took more than thirty years of technological progress in the instrumentation to achieve the sensitivity needed to record this tiny, yet extremely important, ground motion. 
Thanks to these progresses, ``rotational seismology'' was born. This is the field of study devoted to the investigation of all aspects of rotational motions induced by any kind of seismic signal, e.g. local and regional earthquakes, teleseismic signals, cave and mine bursts, environmental noise, volcanic wave field, volcanic activity, etc.\\
Its investigation plays an important role in a wide range of disciplines, from various branches of seismology to earthquake engineering and seismic exploration. The impact on seismology itself is expected to be large and involving many aspects like seismic tomography, point and finite source inversion, volcano source inversion, scattering phenomena constraints, wave field reconstruction, tilt-translation coupling.As a matter of fact, results of modern seismology are still primarily based on the observation of translational ground motions and strain. 
However, since at least two centuries, it was possible to proof theoretically the existence of three components of rotational ground motions around three orthogonal axes. The lack of an adequate broadband instrument delayed their observation until today.
The most reliable instruments to capture rotational motions are optical gyroscopes. The RLG provides top sensitivity and is able to detect any M6+ earthquake. The new portable Fiber Optic Gyroscopes (FOGs), specially designed for seismology, offer the great advantage of portability, which is a mandatory requirement for in field measurements. However, portability entails a lower sensitivity. Since at least ten years, three translational ground motions and one vertical rotation rate are provided by the G-ring in Wettzell, Germany, and since 2014 by the GINGERINO RLG located in the LNGS. Those data permitted to establish the importance of co-located rotational and translational measurements for the study of earthquakes and ocean-generated noise. Still today there is need of conforming and expanding the observations to different sites, possibly in a different structural context and exploring higher amplitude signals and closer epicentral distances with a large RLG. \\
Therefore, we expect very soon advances in this research field thanks to the recent developments in instrumentation and processing techniques.

The GINGERINO experiment started data collection in 2015 and some preliminary papers in the strictly seismological field have been published\cite{GA1,GA2,GA3,AA20,GA5}.

In one paper we present the analysis of rotational and translational ground motions from earthquakes recorded during October–November 2016, in association with the Central Italy seismic sequence. We use co-located measurements of the vertical ground rotation rate from the GINGERINO RLG and the three components of ground velocity from a broad-band seismometer of INGV National Seismic Network. Both instruments are positioned in a deep underground environment, within the LNGS of INFN. We collected dozens of events spanning the 3.5–5.9 magnitude range and epicentral distances between 30 and 70 km. 
This data set constitutes an unprecedented observation of the vertical rotational motions associated with an intense seismic sequence at local distance. Under the plane-wave approximation we process the data set in order to get an experimental estimation of the events backazimuth. Peak values of rotation rate (PRR) and horizontal acceleration (PGA) are markedly correlated, according to a scaling constant which is consistent with previous measurements from different earthquake sequences.
We used a prediction model in use for Italy to calculate the expected PGA at the recording site, obtaining consequently predictions for PRR. Within the modelling uncertainties, predicted rotations are consistent with the observed ones, suggesting the possibility of establishing specific attenuation models for ground rotations, like the scaling of peak velocity and peak acceleration in empirical ground-motion prediction relationships. In a second step, after identifying the direction of the incoming wavefield, we extract phase-velocity data using the spectral ratio of the translational and rotational components. This analysis is performed over time windows associated with the P-coda, S-coda and Lg phase. Results are consistent with independent estimates of shear wave velocities in the shallow crust of the Central Apennines.
In fall 2018, we have observed, by means of an underground ring laser gyroscope, the microseismic (MS) activity induced by the Mediterranean Sea in the 2–5 s period range. In the amplitude spectra, a clear peak highly correlated with a co-located seismometer appears during a large storm activity ranging from 24 November to 1 December 2018. The peak of the spectra, both for rotation rate and acceleration, are found at 3.3 s. We perform the source direction estimation by maximizing the correlation of the vertical rotation rate with the transverse acceleration. \\
In the last paper, we report a study on a set of twenty-two earthquakes that occurred in Central Italy between 2019 and 2020, where we explore the possibility to locate the hypocenter of local events by using a RLG and observing the vertical ground rotation by a standard broadband seismometer. A picking algorithm exploiting the four components (4C) polarization properties of the wavefield is used to identify the first shear onset transversely polarized (SH). The wavefield direction is estimated by correlation between the vertical rotation rate and the transverse acceleration.
The picked times for Pg and Sg onsets are compared to the ones obtained after manual revision on the GIGS station seismometer and are compared with the location provided by the national monitoring service of the INGV. The results obtained suggests values of vp and vs consistent with literature and with the estimation of vs obtained by previous works on this station. The largest part of the events are located in an area that includes the more accurate and precise location of the INGV. By the time of writing this paper, more than 200 earthquakes of magnitude larger than 2.0 occurred from February 2020 in a range of 50 km. The development of an automatic processing routine will permit a more detailed analysis and a deeper learning process for our station with the perspective of future applications in different and less instrumented environments.

\subsection{Geodesy}
The ability to navigate has always been fundamental to the human experience. In the 21st century, more than $ 85\%$ of the world's population (e.g. https://earthweb.com/smartphone-statistics/) has access to smartphones and hence to accurate point positioning. Although services based on positioning are rather easy and accessible to open communities, there is substantial infrastructure required to provide this functionality. Aside from precise orbit determination of the satellites of the Global Navigation Satellite System (GNSS) constellations, highly accurate measurement of the instantaneous rotation rate (hour angle) and orientation of the Earth in space is required.\\
Historically, navigation was performed using star positions as a reference for this purpose. Today, radio telescopes, observing quasar positions at the perimeter of the observable universe, provide us with a long term stable celestial frame of reference. This technique, termed VLBI (Very Long Baseline Interferometry), is used to monitor the Earth Orientation Parameters (EOP) with respect to the inertial reference frame defined by the observed quasars. This reference frame in which both, the quasars and the orbits of the GNSS satellites are defined, is tied to an Earth-fixed, terrestrial reference frame via exact knowledge of the instantaneous rate of rotation of the Earth 
and the orientation of the rotational axis in space (EOP). This terrestrial reference frame, in which we move with a velocity in excess of 300 m/s at mid-latitudes, is what is so important to global society, trade and well being.
Further, it must be known with extremely high accuracy and resolution, since a high reference frame quality allows for more accurate applications whilst errors will quickly grow as the hour angle is integrated up day by day. \\
The Earth system is characterized by the complex interaction of many dynamical subsystems, such as the momentum exchange between the major fluids (atmosphere, oceans, ice shields) and the solid Earth. Mass transport delivered by these fluids on a global scale changes the orientation of the rotational axis and the rotational velocity of the Earth. While variations of these quantities reflect large scale processes, the complexity of these interactions also mean that neither, the effective rotation rate of the Earth, nor the orientation of this vector in the celestial frame of reference can be predicted with sufficient accuracy. Therefore, the rotational velocity of the Earth and polar motion have to be observed on a regular basis. VLBI data, in combination with GNSS observations, are used for this purpose.
Although accessible in principle, VLBI is not directly linked to the rotational axis of the Earth, but to a network of widely spaced radio telescope positions. With the help of a theoretical nutation model, the instantaneous orientation of the Earth rotation axis relative to the body of the Earth can be inferred. In order to satisfy the needs of modern society, we require nine orders of magnitude of resolution for the angular velocity of the Earth, which amounts to a resolution of $50 - 100$ $\mu$s in the length of day and a similarly demanding sensitivity for the orientation of the Earth in space. Aircrafts (for example) are able to navigate autonomously from one airport
to another with the help of accelerometers and inertial rotation sensors. By integrating their sensor signals up, we can obtain the path travelled. In order to apply the same concept of inertial rotation sensing to the Earth, we need gyroscopes that are six orders of magnitude more sensitive than aircraft gyros. Even more challenging is the fact that the sensor stability must also be six orders of magnitude better than that of the aircraft gyros.
Thus, RLGs, and eventually arrays of RLGs, are very promising tools to provide new geophysical data and  to define the relationship among different reference frames.

\section{Experimental set up}
\subsection{Basic design choices}
GINGER is an array of RLGs,  which, as already mentioned, are lasers with a square optical cavity closed by four high reflectivity mirrors\cite{nove,dieci}. Rather common are RLGs with triangular optical cavity, the advantages being that minimizing the number of mirrors, cavity losses are minimized, and that beam paths are always contained in a plane. However, for a given scale factor, directly affecting sensitivity,  more space is needed with respect to square cavities, relevant for underground installation where space is precious. In addition, in the  square cavity configuration  the two diagonals are independent  linear optical cavities, which can be used for fine calibration and geometry control. The angular rotation vector has 3 components, accordingly 3 RLGs are the minimum for a complete investigation. The two RLG array, called GINGER-0, is taken into account as a meaningful step in case of reduced budget. In general, the more RLGs operate within the array, the better the measurement is, since redundancy is valuable for analysis and debugging of the system. \\
It is worth noticing that to make GINGER a multi-site experiment is feasible, being the apparatus of reduced size and costs. In this way the gravitational theories would be tested as a function of latitude, and at the same time systematics of the measurements could be carefully investigated. A multi-site development is of relevance also for geophysics, to better identify local from global signals. Moreover, the mixture of the two fields of science has an added value for the deep comprehension of the apparatus and to constantly compare data with real signals, independently measured and studied.\\
The RLGs of the GINGER array are configured as active Sagnac gyroscopes. However, passive Sagnac gyroscopes, where light is injected into the cavity using an external laser source, have been already developed and tested. Therefore, GINGER could be operated either as active or passive apparatus. To transform GINGER in a passive array of RLGs, we have to remove the active medium (He-Ne gas) and excite the cavities with external laser sources. This possible setup represents a further method to investigate systematics of the apparatus.\\
We also remark that other RLG installations are in operation in Germany and in China\cite{please_Ulli,romy2,JZ2}, that opens the possibility of comparing results from different apparata. Within this frame, on spring 2022 an MoU has been signed between INFN, the Technische Universitaet Muenchen (TUM),  the Ludwig Maximilians Universitaet (LMU), and Huazhong University of Science and Technology (HUST).

\subsection{Sensitivity, design requirements, and main specifications of the apparatus}
RLGs based on square cavities around 4 m in side have already demonstrated able to measure, with a duty cycle close to unity, signals of the order of 1 prad/s in 1 s measurements. As a consequence, GINGER will be perfectly suitable for seismic waves investigations, which requires continuous monitoring with sensitivity better than the nrad/s in the frequency window 0.01-30 Hz. GINGER will fulfill this task as soon as the RLGs, including data acquisition (DAQ) and signal distribution systems, will be operational. Further to that, the plan is to push the sensitivity of the instrument, that is equivalent to say to push the comprehension of this peculiar apparatus, combining optics, atom physics and mechanics, and utilize at best all available signals. Pushing instrumental sensitivity and improving comprehension of the instrumental operation are tasks fully compatible with the continued exploitation of GINGER data in seismic surveys.
Equation \ref{uno}, describing the Sagnac effect, indicates that the target is expressed by $(f_s - <f_s>)/<f_s>$, depending on the different terms $A$, $L$, and $\theta$. In general, each parameter should satisfy the same relative precision with respect to the mean value, or, in other terms, their variation should be kept lower or constantly monitored at this level. For each RLG of the array, the geometrical factor which has to be taken into account is the ratio $A/L$: this is an advantage since we have demonstrated that it is possible to find conditions for the mirror positions in which the mechanical stability affects the scale factor at the second order. The changes in orientation, leading to variations in $\theta$, are unavoidable, but we have shown that they can be accounted for by using signals from at least two RLGs with ad-hoc orientation, meaning that a gyroscope, called RLX, oriented at the maximum signal is required. We note that, GINGERINO is a single RLG, accordingly $\theta$ cannot be determined and as a consequence it lacks of the required accuracy. \\
As a consequence of the above discussion, parameters limiting the performance of the array are: (i) error in the position of the beam spots on the mirrors with respect to the ideal square (this can be quantified by $X0$, the maximum distance of the beam spots center on the mirrors with respect to the vertices of the ideal square);  (ii) the angle between the RLX area vector and the axis of rotation ($\theta0$). We have shown how those two quantities will affect the final result at the second order. \\
The unprecedented high sensitivity planned for GINGER will be achieved by putting together many design strategies, all based on the past activity carried out by the group in the field of high sensitivity RLGs, including realization and operation of prototypes (GINGERINO, GP2, both still in operation) and $R\&D$ projects, such as the G-GranSasso experiment, which in 10 years experimental work has reached meaningful advancements regarding  construction\cite{AA2, AA31}, operation\cite{AA20, AA24, AA33}, control\cite{AA3, AA15, AA12} and laser dynamic comprehension and RLG analysis\cite{AA6, AA11, AA27, AA9, AA10, AA24, AA28} in general. 

\subsection{Single RLG and RLG arrays}
A RLG is a single instrument which is perfectly working as soon as the laser is activated, however the combination of the different RLG of the array is the relevant point to fulfill the targeted precision, and requires to precisely determine orientation of each RLG of the array. 
\begin{figure}
    \centering
    \includegraphics[scale=0.7]{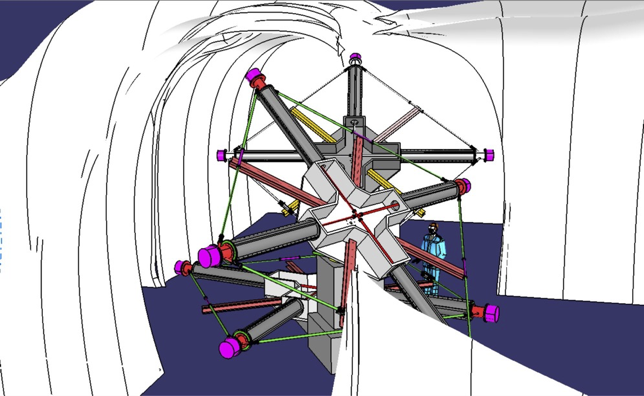}
    \caption{Sketch of the GINGER array assuming co-location of the three RLGs within the Node B at LNGS. The two RLG rings in front will be oriented one horizontally and the other one at the maximum signal (RLX in the following). The third ring has to be out of the meridian plane. This particular relative orientation of the array enables identifying the orientation of the RLGs denoted as RLH and RLV with respect to the axis of rotation. }
    \label{fig:sketch}
\end{figure}
Figure \ref{fig:sketch} shows the sketch of the GINGER array assuming co-location of the three RLGs within the Node B at LNGS. The two RLG rings in front will be oriented one horizontally and the other one at the maximum signal (RLX in the following). The third ring has to be out of the meridian plane. This particular relative orientation of the array allows identifying orientation of the gyroscopes named RLH and RLV with respect to the axis of rotation. Within this approach, RLX is necessary to recover the orientation of RLH and RLV.

\section{Construction and operation}
The construction consists on the assembling of the different parts composing the apparatus, mirror installation, alignment and any operation necessary to make operational the laser. This task has to be carried out by researchers experienced in optics, in this case the two groups of Pisa and Naples; certainly the University of Pisa has the larger experience, as a consequence it has to coordinate the construction. Engineers will be involved as well in the effective assembling procedure, further to contributing to the design development and participating to the meetings with the private companies involved in the manufacturing. 
In the second phase of the project, the operation will be remotely done using the tools developed for GINGERINO and GP2: this will give full access to the experiment operation also for researchers without deep experience in optics.
\subsection{Temporal evolution}
The experiment development, to be started as soon as the RLGs will be in operation, is planned to achieve the following sequential sensitivity steps, where the sensitivity is expressed both as a fraction of the Earth kinetic term and in rad/s; the corresponding requirements in terms of  $X0$ and $\theta 0$ are also reported:
\begin{itemize}
\item	step 0, 1 part in $10^9$ ($7\times 10^{-13}$ rad/s);  ($X0 < 50\mu$m, $\theta0 < 50\mu$rad)
\item	step 1, 1 part in $10^{10}$ ($7\times10^{-14}$ rad/s); ($X0 < 16\mu$m, $\theta 0 < 16\mu$ad)
\item	step 2, 1 part in $10^{11}$ ($7\times10^{-15}$ rad/s); ($X0 < 5\mu$m, $\theta 0 < 5\mu$rad), this amount to the $10\%$ test of gravitomagnetic effects.
\end{itemize}
The planned steps will start already at the end of the construction phase and will be scheduled by allocating a sufficient time for both the RLX alignment procedure, and the subsequent assessment of the achieved sensitivity enhancement, based on data taking and analysis. While data analysis for the single ring signal is well established and reliable, gaining insight into the set of the array signals has been never performed at this level of sensitivity. To this end, novel data analysis routines will be implemented. Sensitivity depends on the alignment of RLX with the rotation axis, and the target of  $10\%$ test of Earth gravitomagnetic effects will require to align RLX at the level of $2$ $\mu$rad. \\
It is worth noticing that each alignment step will provide useful data both for fundamental studies and geoscience. As a matter of fact, fundamental physics effects depend on latitude and they  have never been measured ``locally'' on the Earth surface. Moreover, and each step will improve the sensitivity of the LV tests as well, as it pushes forward the constraints on some families of extended theories of gravity. In particular, a sensitivity of 1 part in $10^9$ of the Earth rotation rate is the first target to get experimental evidence for the GR terms and to provide limits for LV parameters that are competitive with existing results. \\
We remark that the Earth global signals are measured with very high accuracy by IERS, each day the Length of Day (LoD) error is 3-4 $\mu$s. They will be measured by GINGER as well, with a clearly completely different approach enabling Earth-based measurements and adding information on the fast variation of LoD. IERS data can therefore be effectively used as reference signals for fundamental physics and geophysics investigation. \\
At the end of this 4 year project it will be clear whether GINGER can be further used to extend the sensitivity limits. GINGER is based on a modular structure (Hetero Lithic, HL), which forms the structure of the optical cavity. The long-term response will be affected by this structure. The mechanical structure of the optical cavity is certainly the most important part of the apparatus. The first requirement is rigidity and  to have proper resonances of the structure above 50 Hz, meaning that no structural resonance should appear where the Sagnac frequency is expected. All the above requirements have been duly considered in the design. Since the expected signals require long data taking, the scale factor, proportional to the area over perimeter ratio, has to be stable. As a consequence the long term stability and any kind of deformation of the cavity are points of concern. The planned tests and the engineering studies will be relevant to follow and keep under control the geometry using the optical cavities along the diagonals as a monitor.  At the end of the project limitations of the setup will be clear, and retrofitting design will indicate the improvements required to further improve sensitivity.


\subsection{The problem of the shot noise}

Any measurement based on the detection of photons is limited by an intrinsic limit: the shot--noise level (SNL). This limit comes from the quantum nature of light and is often referred as standard quantum limit (SQL). The inherent fluctuations in the number of photons that stocastically are converted in photo-electrons set an intrinsic limit to the precision of the measurement. This limit depends on the effective description of the measurement in a quantum model. While rarely SQL becomes an effective limit to the precision one can achieve in an optical apparatus, it is well known that it limits interferometric measurements \cite{Caves81, chow} and methods exist to overcome it \cite{LIGO}.

Theoretical estimation of SQL in RLGs, where two beams of slight different frequency interfere, have been so far obtained under the strong assumption of two completely uncoupled optical fields oppositely propagating inside the ring cavity \cite{Cresser82}. The above assumption does not take into account the unavoidable coupling of the two beams that, in a fully classical picture, manifest as back-scattering \cite{Wilkinson87}.

Being our device one of the most sensitive one, we believe that it is the time to  re-write quantum fluctuations equations of the relevant operators in more complete model so to have a better comprehension of the physical mechanisms fixing its ultimate noise limit.

\section{Summary of the main experimental aspects}

The most relevant design requirements can be summarized as follows.
\begin{itemize}
\item 	Prerequisite for achieving the above mentioned sensitivities is a sufficiently large scale factor, ($SF=4A $/$ \lambda L$). Since it ultimately depends on the size of the RLG optical cavity, square rings with a minimum side length of 4 m are required. Final definition of the size will obviously depend on the availability of space in the installation site. Note that a modular design will be followed, enabling realization of shorter length prototypes (GINGER-0) in the initial stage of the project, if required, aimed at defining the experimental procedures for the RLG alignment.
\item Aim of the experiment is to reconstruct the angular velocity vector. Therefore, an array of independent and co-located RLGs will be realized. A minimum of two RLGs have been demonstrated sufficient to retrieve the main features of the angular velocity vector, although full three-dimensional reconstructions would need three RLGs, that would also be beneficial in terms of redundancy. The modular approach will allow us to tailor the design also according to external constraints on the availability of space in the installation site.
\item Sensitivity depends on the precision achieved in calibrating the scale factor $SF$. The main difficulty for fundamental physics tests is the subtraction of the kinetic term of the rotating Earth, which is $10^9$ times higher than the signals we are looking for, and for this reason any relative misalignment of RLGs or fluctuations in their scale factor, changing the projection of the dominant term, must be carefully estimated in order not to bias the gravito-electric or gravito-magnetic signals. Strict design requirements on the mechanics of the instrument and an original layout, including materials, will mitigate the occurrence of fluctuations, fully benefiting of the inherent advantages (low anthropic noise, stable environmental conditions) of the underground location, already widely demonstrated by the GINGERINO prototype. 
\item An additional design requirement is the possibility to implement active controls of the optical cavity geometry, based on interferometric techniques. For example a sensitivity of 1 part in $10^9-10^{11}$  of the Earth rotation rate imposes the stabilization of the dimensions of the optical cavity at the level of 1 part in $10^9-10^{11}$  relative to its sides. If we consider an RLG with a side of few meters, this requirement implies to measure all the lengths with an accuracy in the nanometer to picometers range. However, we have shown that when the ring cavity shape, is close to an ideal square, the $SF$ depends quadratically on the deformations of the optical cavity, and the requirements for the measurement of cavity sides are in the range of fractions of millimeters down to fractions of micrometers. The instrumental design includes both the possibility to continuously monitor the geometry via Fabry-Perot cavities along the diagonals of the square cavity and the ability to electrically actuate the cavity mirror holders. Feedback techniques, already established and tested in the GP2 prototype, will enable locking the cavity size down to the micrometer, or even fraction of micrometers, range, in order to meet the targeted sensitivity level. 
\item A key and distinctive point of GINGER concerns the estimate of the relative angle $\theta$ entering the scale factor $SF$. It has been proposed in the past that such an angle can be deduced by measuring the relative angles between different RLGs, in principle feasible by optical techniques. However, the requirements are in the range of nrad to  prad, and therefore rather demanding and expensive techniques should be implemented. The alternative solution we have proposed and studied in 2017 is based on orienting one of the RLGs of the array (denoted as RLX) at the maXimum of the Sagnac beating frequency, i.e. with its area vector parallel to the angular rotation axis; RLX measures the absolute value of the total angular rotation and it is affected by tilts at the second order, hence it can be used to estimate the angle $\theta$ of the other RLGs of the array. In this case, the requirement on the absolute orientation of RLX with respect to the total angular rotation vector is $50-0.5$ $\mu$rad, depending on the sensitivity level. A key design parameter is therefore to realize a specific mechanical arrangement enabling controlled rotation in the aforementioned range of RLX. This will allow us to avoid an external metrology system while reaching the sensitivity limit set by the RLGs noise, according to the above declared sequential steps. In the initial phase of the construction, an alignment precision in the order of the mrad will be sufficient for the RLX. However, to increase sensitivity we must perform on-site a fine adjustment of its orientation, as allowed by the proposed design where RLX can be  tilted as a whole with respect the North-South and East-West directions.  Since this operation has never been done before, and relevant experimental procedures are missing, the GINGER experiment will proceed by steps with the aim of reaching the final goal after several iterations, each one validated via data acquisition for a sufficiently long time. 
\item The proposed 3 RLG array is able to recover the signals for the 3 dimensions investigation. The addition of other RLGs would be advantageous by providing redundancy; the modularity of the proposed scheme allows expansion of the array adding more RLGs. The RLG array composed of RLX and RLH is the minimum setup to reconstruct the orientation with respect to the local rotational axis. While not powerful like a full three RLG array, it is still meaningful to reconstruct the orientation of RLH, which has never been done so far. The third RLG, denoted as RLV, could be installed at a later stage.
\item Sensitivity is the key point to reach the fundamental science goals, and the size of the square cavity rules the sensitivity. Since the most important signals are at low frequency, stability of the apparatus is the other key point, and the material used for the RLG structure ultimately rules its stability. Certainly, too large setups would not be stable in the long term, so sensitivity and stability have to go together. GINGER is the evolution of the GINGERINO experience with a RLG setup linked to a single structure rigidly attached to the ground in the central point, and contained in a thermal bath to avoid perturbations. Considering the space available in Node B, it is possible to accommodate 3 RLGs with side from 4 to 6 m, keeping in principle room for a fourth RLG. The most effective choice is to have the RLGs with a side length of maximum 5 m, since mirrors of 1” diameter and 4 m radius of curvature can be used; such mirrors are similar to the ones adopted for GINGERINO, and the chosen side length leaves enough room for the installation. 
\item The long term stability depends on the material used for the HL structure. Two materials have been selected: granite and silicon carbide. Granite has been already tested for our purposes, GP2 and GINGERINO are mounted on top of granite tables. In the design structure proposed for GINGER, called GP3, each single RLG has a central structure in granite, and spacers to bring the side from 2 up to 6 m length. Moreover,  vacuum tanks will be made in titanium, in order to reduce the weight of the vacuum system, which has to be positioned at the tip of the spacers. For the spacers, assuming a 5 m side length, the best choice is to use silicon carbide bars.
Silicon carbide is a ceramic material having better properties with respect to granite, in terms of Young modulus and thermal expansion coefficient. The use of granite spacers is a cheap solution for RLGs, in this case the maximum side length being 4 m. The close collaboration with the engineers will allow the precise modeling of the apparatus and its test, to study and improve the proposed RLG HL structure. The role of deformative behaviour in the functioning of RLG has been recognised to be extremely relevant. A short list of the possible phenomena which may influence the long term response and sensitivity of the apparatus is: (i) inhomogeneity and anisotropy of the candidate materials (granite or silicon carbide) to be used for the body of the spacers and the whole structure; (ii) the creep and plasticity phenomena in an apparatus having heavy hanging structural elements; (iii) the imposed pre-stresses and pre-deformations induced by changes of environmental temperature, humidity and permanent damage inside structural elements; (iv) friction and dissipation phenomena induced by accidental dynamic actions; (v) the induced deformation in the whole RLG due to the unavoidable presence of surfaces of discontinuity of the mechanical properties at the junctions between different structural elements. The non-standard structure which has been conceived and the extremely high precision of the demanded measurements oblige the designer to modify the classical paradigm in the conception and control of the structure, by including the control of effects usually considered negligible. The consequent analysis will require the development of novel continuum mechanical models and non-standard numerical codes to get a reliable prediction of the long term behaviour of the structure. Already available standard design techniques allowed us to conceive an apparatus which surely has the minimal sensitivity performances needed to ensure the success of the project. However, we expect that, for getting the maximum possible sensitivity obtainable for RLG, the micro-morphic higher gradient continuum models will be needed in order to obtain the detailed predictivity imposed by the considered structural problem: in this case novel iso-geometric numerical integration techniques will be required together with innovative algorithmic solution schemes. In this context, retrofitting design is essential. Statistical analyses will be carried out on properly sampled tomographic pictures to experimentally characterize the microstructure of the employed materials, including anisotropy, porosity, etc. In situ X-ray tomographic measurements of granite and silicon carbide specimens subjected to various mono-axial and multi-axial loads will be carried out along with force measurements to optimally characterize the constitutive parameters of the above-mentioned micromorphic continuum modeling. Thermographic measurements will help towards the understanding of possible dissipation phenomena and of the emergence of temperature-induced non-uniform spurious deformations.

	\end{itemize}
\section*{Acknowledgments}
We warmly thank Angelo Tartaglia for his constant scientific support and stimulating discussions. The $R\&D$ activity toward GINGER has been financed by the INFN Astroparticles Commission (CSN2).
  

\

 \end{document}